\documentclass[allclo,superscriptaddress,eqsecnum,amsfonts,showpacs]{revtex4}
\usepackage{epsfig}

\newcommand{\be}{\begin{equation}}
\newcommand{\ee}{\end{equation}}
\newcommand{\bea}{\begin{eqnarray}}
\newcommand{\eea}{\end{eqnarray}}

\newcommand{\ep}{i\varepsilon}
\newcommand{\nn}{\nonumber}

\begin{document}

\preprint{ \parbox{1.5in}{\leftline{hep-th/??????}}}

\title{Quark Schwinger-Dyson equation in temporal Euclidean space}

\author{V.~\v{S}auli}
\affiliation{Dept. of Theor. Phys., INP, \v{R}e\v{z} near Prague, AV\v{C}R}
\affiliation{CFTP and Dept. of Phys.,
IST, Av. Rovisco Pais, 1049-001 Lisbon,
Portugal }
\author{ Z. Batiz}
\affiliation{CFTP and Dept. of Phys.,
IST, Av. Rovisco Pais, 1049-001 Lisbon,
Portugal }

\begin{abstract}
We present an elementary nonperturbative method to  obtain Green's functions (GFs) for timelike momenta. We assume there are no singularities in the second and the fourth quadrants of the complex plane of space momentum components and perform a 3d analogue of Wick rotation. This procedure defines Greens functions in a timelike Euclidean space. As an example we consider the quark propagator in QCD. While for weak coupling, this method is  obviously equivalent to perturbation theory, for a realistic QCD coupling a complex part of the quark mass and renormalization wave function has been spontaneously generated even below the standard perturbative threshold. Therefore, our method favors a confinement mechanism based on the lack of real poles.

\end{abstract}

\maketitle
%

\section{Introduction}

The observable spectra of hadrons represent clear information about the S-matrix at  color singlet channels.
The processes $e^-e^+\rightarrow hadrons$, 
$\tau $ hadronic decays and so on, are significant sources of experimental information in timelike regime of momenta. The microscopic description of such phenomena at  low energy QCD processes are  intuitively  understood in terms of  elementary QCD quanta, although the quantitative  description is almost missing. Also other quantum field theoretical nonperturbative problems,e.g. confinement, require computation of Green`s function (GF) for timelike momenta. 
For understanding QCD the knowledge of GFs is a crucial matter. As  QCD is commonly accepted as the  ordinary  quantum field theory, the amplitudes  should be obtainable from elementary QCD GFs which could already encode the information about observables.

  Considering the quark propagator in momentum space, the absence of  real poles  is a tempting idea of quark confinement. The spontaneous dynamical generation of an  imaginary part of the quark mass can lead to the absence of real pole.
We argue that this is the scenario of confinement by showing the model where the imaginary part of quark propagator is induced  below the expected perturbative threshold. Recall that the threshold value would be otherwise  uniquely determined just by the  quark pole mass. Using the formalism of Schwinger-Dyson equations, we will exhibit a realistic scenario, in which the quark mass function, as well as the quark renormalization function, become complex for almost all timelike momenta. As usually, the quark  propagator remains real for  spacelike momenta, where its values  correspond to the results performed in  the standard Euclidean formalism.

Euclidean space Lattice theory represents the method which  in principle provides the information about GFs from the first principles. Minkowski space simulation in QCD is not recently  feasible because of oscillating phase factor in the generating functional. To make the method feasible, the continuation to the imaginary time axis is required and the problem is solved in unphysical Euclidean space. Afterward the continuation back is necessary. Such a continuation of lattice data to the timelike momentum axis has been performed only very recently \cite{LANES} providing thus data for very low momenta only. Avoiding large systematic errors, the continuation can be performed only within imposing of an additional global analytical  assumption \cite{SAULI}.

In this paper we will solve the quark gap equation which is an alternative way 
to achieve the non-perturbative solution  for QCD Greens functions 
\cite{QCDinfrared,FISCHER2006} . The quark gap equation is the part of the Schwinger-Dyson equations (SDEs) which when solved exactly could provide the fully dressed Greens functions as well. As SDEs are an infinite tower of coupled integral equations, they require approximation and/or  truncation of the SDEs system.
 Similarly to lattice formulation of quantum field theory, the most studies of SDEs are performed in the Euclidean space.  The trial functions with  given analytical properties had been used to make naive continuation to the timelike regime. The results of the paper \cite{Alkac2} point towards an analytical structure of the quark
propagator with a dominant singularity on the real timelike axis, while the nature of this singularity has not been determined with confidence. In the light of very recent numerical study \cite{LANES}, the singularity can be a branch point and not a real or  complex conjugated poles suggested in \cite{Alkac2}. The other studies preformed on various assumption also point towards the absence of real pole or they at least challenge that the real pole could be a dominant singularity of the quark propagator \cite{SAULI2,SAULI3,MARIS}.

Perturbation theory (PT) is the only known method where such continuation is well  understood and massively used in practice. In fact in PT at finite order, the analyticity assumptions are more specified: the propagators are  analytical functions in the whole complex plane  up to a real positive semi axis of $q^2$.
When PT is reliable,   the particles are revealed in the GFs poles 
and branch cuts in momentum space. In such circumstances the tree level single pole propagator is dressed within its form constrained by
the PT analyticity  described above. In  general  it leads to the known forms of integral representations and
 dispersion relations for GFs. This has been used in QCD SDEs formalism long time ago \cite{CORNWALL},
assuming  that the spectral representations remain valid for the full nonperturbative solution.
More recently, the method of solution based on  such spectral representation has been been checked in practice for the number of the  toy models \cite{SAULI3,BLUMA,SAULIJHEP}.
providing the  correct solution only  for rather weak coupling, while it appears to be  inefficient when the  couplings  exceed certain critical values.  It is to be noted, that the position of the branch points are uniquely dictated by the mentioned spectral  method, which gives us a little freedom for spontaneous generation of complex masses. The above mentioned facts do not disprove the spectral method completely, however  the  practical failure of the method could be understood as a sign of weakness of the analytical assumptions.

In the next Section  we propose new approach based on weaker analytical assumption
(compared to the PT or spectral technique discussed above) and  3d Wick rotation is introduced to "rotate" originally space components to imaginary axis.
The method  is applied to the quark gap equation in the Section 3. 
The obtained  solutions is presented in  Section 4.

\section{From Minkowski space to temporal Euclidean space $E_T$}

In lattice theory and in most of the Schwinger-Dyson equations approaches in the literature the so called Wick rotation is used to avoid calculations in Minkowski space, wherein the Green`s functions are singular and the integration is problematic, especially numerically.
Besides the singularities problem, there is another aspect impeding momentum integration presented in SDEs, the hyperbolic angle of Minkowskian "spherical" coordinates:
\bea
&&k_0=k \cosh\theta_1
\nn \\
&&k_x=k \sinh\theta_1 \cos\theta_2
\nn \\
&&k_y=k \sinh\theta_1 \sin\theta_2 \cos\theta_3
\nn \\
&&k_z=k \sinh\theta_1 \sin\theta_2 \sin\theta_3
\eea
runs from - $\infty$ to $\infty$ and most (principal valued here) integrals in terms of these
hyperbolic angles cannot be found in a closed form. (To that point, there exist a semiperturbative prospect to work directly in Minkowski space,  the first iteration of SDE was performed in the paper \cite{PEDRO}, however single real pole propagator was necessary input to perform some integration analytically)

The mentioned Euclidean space formulation, however, has some drawbacks, since physics involves Minkowski space and not Euclidean one. One needs to "rotate back" the results to obtain them for timelike arguments. This rotation is basically an analytical continuation on the boundary of perturbative analyticity domain and therefore can be frequently ambiguous, especially if the results are numerically obtained.

In order to circumvent  the difficulties stemming from the Minkowski metric or inverted rotation we propose a different procedure: instead of Wick rotating the time variable we rotate the space components.
Clearly, this way we maintain the singularity structure, which for a free propagators stays on the exterior boundary of complex contour, but this is a small price to pay for the fact that angular integral are more tractable and especially for the fact that we do not need to rotate the variable twice. The method we will present here is used to obtain results at timelike regime of fourmomenta, while the correlation functions for spacelike arguments can be evaluated in the standard fashion.

\begin{figure}
\centerline{\epsfig{figure=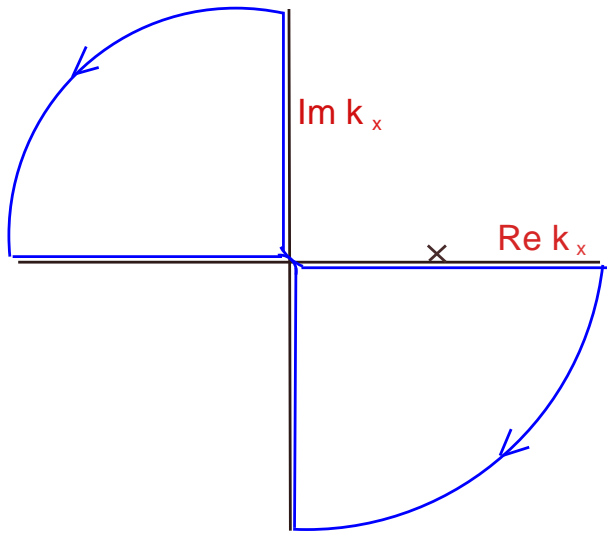,width=8truecm,height=7truecm,angle=0}}
\caption[caption]{The Figure shows the integration contour in complex plane for the first space component of the momentum. The perturbative singularities (cross) lies in exterior vicinity of the contour. } \label{wrau}
\end{figure}

We assume there are no singularities in the second and the fourth quadrants of complex planes of the complex variables $k_x,k_y,k_z$. Giving the Lorentz invariance, the singularities in the kernels can be  functions only of $p^2$, this assumption is in agreement with the  one used in standard Wick rotation, wherein there are no assumed singularities in the first and the third quadrants of the complex $k_0$ plane. This happen for instance when the obtained imaginary part of the square of the mass function is negative, excluding thus any singularities from the I. and III. quadrants and the imaginary $k_i$ axis  as well.   The afore-mentioned Wick rotation is sketched in Fig. \ref{wrau}. Cauchy theorem gives the following prescription for momentum:
\bea
&&k_{x,y,z} \rightarrow i k_{1,2,3} \, ,
\nn \\
&&i \int d^4k \rightarrow  \int d^4k_{E_T} \, ,
\eea
 which in the case of original $3+1$ is identical with standard  Euclidean $E$ "spacelike" one. Note only that, the additional $i$ appears when the original Minkowski space is of odd-dimensionality. 

For instance the free propagator of scalar particle then looks
\be
\frac{1}{p^{2}-m^2+\ep} \, ,
\ee 
with positive square 
\be
p^2=p_1^2+p_2^2+p_3^2+p_4^2.
\ee

If necessary, one can use  the Euclidean definition of Dirac gamma matrices,  
$\gamma_o\rightarrow \gamma_4 \, ; \vec{\gamma}\rightarrow i\vec{\gamma}_{E_T} $
and redefined gamma matrices satisfy 
$\{{\gamma}^{\mu}_{E_T},{\gamma}^{\nu}_{E_T}\}=2\delta^{\mu\nu}$.

Since the  fixed square Minkowski momentum $p^2=const$ hyperboloid with infinite surface is transformed into the finite four-dimensional sphere in $E_T$ space, the Cartesian variable are related to the  spherical coordinates as usually:
\bea
&&k_4=k \cos\theta
\nn \\
&&k_1=k \sin\theta \cos\beta
\nn \\
&&k_2=k \sin\theta \sin\beta \cos\phi
\nn \\
&&k_3=k \sin\theta \sin\beta \sin\phi \, .
\eea 

\section{Quark SDE}

In QCD the quark propagator $S$ is conventionally  characterized by two independent scalars, the mass function $M$ and renormalization wave function $Z$ such that
\be
S(p)=\frac{Z(p)}{\not p-M(p)+\ep}=[\not p A(p^2)-B(p^2)+\ep]^{-1}\, ,
\ee
or equivalently the functions $A,B$ (where simply $M=B/A$ , $A=1/Z$) are used when suitable,
 noting for the bare fermion propagator $S_0$ we have $A=1$ and $B=m_0$.
If the function $M(p)$ preserves the real pole in the full propagator $S(p)$ then the insertion of Feynman $\ep$ defines the way of loop momentum integration, otherwise it can omitted.  
For shorthand notation, we will also express the quark propagator in terms of the Dirac vector ($S_v$) and the Dirac scalar ($S_s$) parts of the propagator: 
\be
S(p)=\not p S_v(p^2)-S_s(p^2).
\ee

The gap equation for the inverse of $S$ is
\bea \label{zdar}
S^{-1}(p)&=&S^{-1}_0(p)-\Sigma(p) \, ,
\nn \\
\Sigma(p)&=&i C_A g^2\int\frac{d^4q}{(2\pi)^4}\Gamma_{\alpha}(q,p)
G^{\alpha\beta}(p-q)S(q)\gamma_{\beta} \, ,
\eea
where  $\Gamma$ is the quark-gluon vertex, 
 $C_A=T_a T_a=4/3$ for $SU(3)$ group and $G^{\alpha\beta}$ is the gluon propagator, which in  covariant  gauges reads

\be  \label{gluonprop}
G^{\mu\nu}(k)= 
\left[-g^{\mu\nu}+\frac{k^{\mu}k^{\nu}}{k^2}\right]G(k^2)
-\xi \frac{k^{\mu}k^{\nu}}{({k^2})^2}  \, ,
\ee
where $G$ at tree level reads
\be
G(k^2)=k^{-2} \, ,
\ee

 The gauge parameter dependent term in Eq. (\ref{gluonprop}) remains undressed unless the gauge symmetry is  broken, which we assume is the case of QCD.

As QCD is a non-Abelian gauge theory, the GFs are essentially gauge dependent, however in any gauge, the various Greens functions are related through the complicated Slavnov-Taylor identities. These constraints are especially simplified in the Background Field Gauge (BFG), wherein they simplify to the sort of Ward-Takahashi identities \cite{BFG1,BFG2,BFG3,BFG4}. In this case, the symmetry of the system has not been changed by gauge fixing procedure. 
 When solving Schwinger-Dyson equation the BFG  can be usefully further exploited 
\cite{BFG4,BIPA2008,ABP08}, even without the exact knowledge of the missing and unknown vertices.

Especially, in BFG the quark-antiquark-gluon-vertex function satisfies QED like  Ward identity (WTI):
\be \label{WTI}
k^{\alpha}\Gamma_{\alpha}(p,l)=S^{-1}(p)-S^{-1}(l)\, ,
\ee
where $k=p-l$ is the gluon fourmomentum. We use the advantage of BFG and  we will consider quark propagator in BFG in this paper.

\section{Metric tensor truncation of quark SDE}

In this section we transform the Minkowski quark SDE into the 
two-dimensional equation in Euclidean temporal space. To proceed this we first specify  the  approximation of the SDE.
In order to go step further beyond the simplest ladder approximation  we will use the exact WTI of BFG to treat product $k \cdot \Gamma$ in the kernel of SDE (\ref{zdar}). This allows  to entirely evaluate the contribution wich stem from this term.

The remaining what needs to be specified is the product of the full vertex  $\Gamma$ with  the metric tensor  part $g$ of the gluon propagator . As an introductory approximation made in the temporal Euclidean space, we simply take for the product $g_{\mu\nu}\Gamma^{\mu}\approx \gamma_{\nu}$. The approximation is improvable by making a loop expansion with dressed internal propagators (i.e. skeleton expansion). This is a future program which, in addition,  will check the reliability   of approximation used here.

For convenience we will denote 
\be
\Sigma=\Sigma_T+\Sigma_{L\xi}=\Sigma_g+\Sigma_L+\Sigma_{L_\xi}
\ee
\be
\Sigma_{i}(p)= \delta B_i(p) - \delta A_i(p)\not p
\ee
for $i=g,L,L_\xi$,
where $\Sigma_T$ stems from the dressed transverse part of the gluon propagator and 
 $L,(L_\xi )$ labels the selfenergy contribution which follows from the dressed (undressed gauge) longitudinal term in gluon propagator, clearly $T=g+L$ in our notation. 

In this notation the appropriate terms explicitly read
\bea
\Sigma_{g}(p)&=&-iZ_1g^2C_A \int_k \Gamma_{\mu}(k,p)g^{\mu\nu}G(q)S(k)\gamma^{\nu} \, ,
  \\
\Sigma_{L}(p)&=&iZ_1g^2C_A \int_k \Gamma_{\mu}(k,p)\frac{q^{\mu}q^{\nu}}{q^2}G(q)S(k)\gamma^{\nu} \, ,
 \\
\Sigma_{L\xi}&=&-iZ_1g^2C_A\xi \int_k \Gamma_{\mu}(k,p)\frac{q^{\mu}q^{\nu}}{({q^2})^2}S(k)\gamma^{\nu} \, , \label{gaugeterm}
\eea
where we have used the shorthand notation $\int_k$ for the fourdimensional  integral $\int\frac{d^4k}{(2\pi)^4}$.

Using the WTI we can  get for   $\Sigma_{L\xi}$  
\bea \label{boruvka}
\Sigma_{L\xi}(p)&=&-i Z_1 g^2 C_A \xi \int_k \frac{\not k}{({k^2})^2}
\nn \\
&+&i Z_1 g^2 C_A \xi \int_k  S^{-1}(p)S(k)
 \frac{\not q}{(q^2)^2} \, .  
\eea
The first term (\ref{boruvka}) is zero since  it is odd in the variable $k$.

Performing  3d Wick rotation and integrating over the Euclidean angles we get the following contribution to the  renormalization function:
\be
\delta A_{L_\xi}(x)=\frac{\xi g^2 C_A}{(4\pi)^2}
\left[B(x)\int_0^{x} dy \frac{y}{x^2}S_s(y)
+A(x)\int_x^{\infty} dy S_v(y)\right]\, ,
\ee
where $x=p^2_E$ and $y=q^2_E$.
 
For the contribution to the function $B$ we can obtain
\be
\delta B_{L_\xi}(x)=\frac{\xi g^2 C_A}{2(4\pi)^2}
\left[A(x)\int_0^{x} dy \frac{y}{x}S_s(y)
+B(x)\int_x^{\infty} dy S_v(y)\right] \, .
\ee

In order to  calculate $\Sigma_T$ the transverse part of the full gluon propagator
needs to be specified. At low $Q^2$ the BFM gluon propagator  is unknown function of momenta and gauge parameter, the only known  is the undressed longitudinal part. To that point we will consider Landau gauge,  assuming that  various recent studies  SDEs and lattice calculations performed in this gauge, offer already reasonable estimate. The most ambiguous is the deep infrared behaviour $q^<\Lambda^2_{QCD}$, depending on the details, most of the recent studies shows up that tree level $q^2=0$ pole singularity is softened  \cite{Alkac2} $1/(q^2)^a$, $a<1$, with possible infrared finite solution  \cite{AGUNA2004}, \cite{AGBIPA}.

In  the present paper we assume $q^2$ is a branch point of gluon propagator, which does not have a purely real pole in its transverse part. More specifically,  we will assume that the product of the coupling with  gluon propagator can be expressed through the following integral representation:

\be  \label{rep}
\frac{g^2}{4\pi}G(q^2,\Lambda_{QCD})=\int_0^{\infty} d\nu\, \frac{\rho_g(\nu,\Lambda_{QCD})}{q^2-\nu+\ep} \, . 
\ee 
Thus, contrary to studied quark propagator, the standard analyticity for gluon propagator is still assumed. As already mentioned, such a representation has been already used in SDE context \cite{CORNWALL}, however  that the exact gluon propagator  has not the assumed analytical properties is quite possible which would complicate our analysis in this case.

To do our best we will use the  reasonable model of the  gluon propagator at all scales. Below, we discuss several  basic requirements which  should be satisfied.

Firstly, the prescription (\ref{rep}) will respect asymptotic freedom thus for sufficiently large $q^2$ the leading power behaviour  must be softened by standard 
perturbative log corrections such that
\be \label{prop}
\frac{g^2}{4\pi}G(q^2,\Lambda_{QCD}) \simeq \frac{1}{q^2 \, log (q^2/\Lambda_{QCD}^2)+...}
\ee
where the dots represents higher order scheme dependent contribution.

It will have no unphysical singularity (known from naive use of perturbative theory at strong coupling). At last but not at least,  the $\rho_g$ in the gluon propagator may involve confinement. The last two requirements listed above are automatically satisfied for any regular function $\rho_g$. To comply with this we will not assume that $\rho_g$ includes Dirac delta  as it would be when free particle mode is expected.

To satisfy all the  requirements simultaneously
the  propagator function $G$ can be constructed by considering the function 
\be \label{gluonsr}
\rho_g(x)=2\frac{\alpha(x)}{\alpha(0)}\frac{\rho_{\alpha}(x)}{x}
\ee
where the function $\alpha(x)$ is calculated through
\bea
\label{repa}
\rho_{\alpha}(x)&=&\frac{4\pi/\beta}{\pi^2-\ln^2{(x/\Lambda^2_{QCD})}} \, ,
\nn \\
\alpha(x)&=&P. \, \int_0^{\infty} d\nu\, \frac{\rho_{\alpha}(\nu)}{x-\nu} \, ,
\eea
where symbol $P.$ stands for Cauchy principal value integration and  $\beta$ in (\ref{repa}) represents the beta function coefficient, for which we take $4\pi/\beta=1$ (recall,  $4\pi/\beta=1.396$ for three active quarks in perturbative QCD).

Recall also, the auxiliary functions $\rho_g$,  $\alpha(x)$ correspond to the imaginary  and real parts of the analyticized  1-loop effective
charge $\alpha_{QCD}(-x)$ constructed in \cite{AC1,AC2,AC3}, however the original  meaning of $\alpha(x)$ is lost here. In our approach it is  the gluon propagator, and not the running charge, which satisfies dispersion relation (\ref{rep}). The full expression for $\alpha$ can be found in the original paper.


Substituting IR (\ref{rep}) into $\Sigma$ we can write for $\delta A_g$
\bea
\delta A_g(p^2)&=&-\frac{Tr (\not p \Sigma_g(p))}{4 p^2}
\nn \\
&=&-i4\pi C_A \int_k\frac{2p\cdot k}{ p^2} S_v(k^2)\int_0^{\infty}d\nu \frac{\rho_g(\nu)}{q^2-\nu+\ep} \, ,
\eea
where $q=p-k$. Performing the 3d Wick rotation and integrating over the Euclidean angles we get
\be \label{bg1}
\delta A_g(p^2)=\frac{-C_A}{\pi^2}
\int_0^{\infty} dy y \sqrt{y/x} S_v(y)
\int_0^{\infty}d\nu \rho_g(\nu) 
I_2(x,y,\nu) \, ,
\ee
where the function $I_2$ is defined below by (\ref{I2}).
Similarly we can easily derive the contribution from $g$ to the function $B$
\bea \label{bg}
\delta B_g(x)&=&\frac{Tr}{4} \Sigma_g
=\frac{2C_A}{\pi^2}
\int_0^{\infty} dy y  S_s(y)
\int_0^{\infty}d\nu \rho_g(\nu) 
I(x,y,\nu)\, .
\eea

The functions $I,I_2$ in (\ref{bg}) and (\ref{bg1})  are the complex non-holomorphic functions defined through the angular integral in the following
way 
\bea
I(x,y,\nu)&=& \int_0^{2\pi} d\theta \frac{\sin^2\theta}{x+y-\nu -2 \sqrt{xy} \cos \theta +\ep} \, ,
\\
I_2(x,y,\nu)&=&\int_0^{2\pi} \frac{d\theta \sin^2\theta cos\theta }{x+y-\nu -2 \sqrt{xy} \cos \theta +\ep} \label{I2} \, .
\eea
Both integrals above can be evaluated in a closed form and we list the results in the Appendix A.

For $L$ contribution  we first use the WTI (\ref{WTI}) and the integral representation (\ref{rep}), then the appropriate contribution  can be written like
\be \label{sigmal}
\Sigma_L(p^2)=i 4\pi C_A \int_k \frac{\not q}{q^2}
S(k)S^{-1}(p)\int_0^{\infty}\frac{d\nu\rho_g(\nu)}{q^2-\nu+\ep} \, .
\ee
Making the appropriate trace   projections, performing the 3d Wick rotation and after some trivial manipulations we get
\bea
\delta A_L(x)&=&\frac{c_p}{x} \int_0^{\infty} d \nu  
\rho_g(\nu) \int_0^{\infty} dy y 
\left[-S_v(y)A(x)x\int_0^{\pi}\frac{d\theta q.k \sin^2 \theta}{q^2(q^2-\nu+\ep)}
+S_s(y)B(x)\int_0^{\pi}\frac{d\theta q.p \sin^2 \theta}{q^2(q^2-\nu+\ep)}\right]
\, ,
\nn \\
\delta B_L(x)&=&c_p \int_0^{\infty} d \nu  
\rho_g(\nu) \int_0^{\infty} dy y 
\left[-S_v(y)B(x)\int_0^{\pi}\frac{d\theta q \cdot k \sin^2 \theta}{q^2(q^2-\nu+\ep)}
+S_s(y)A(x)\int_0^{\pi}\frac{d\theta q \cdot p \sin^2 \theta}{q^2(q^2-\nu+\ep)}\right]\, ,
\eea
where
\be
c_p=\frac{4\pi C_A}{(2\pi)^3 } \, .
\ee

In the above formula we do not state explicitly the fact that the all scalar products are  in $E_T$ space. The scalar products $k\cdot q=k^2-k \cdot p $ and $q\cdot p=k \cdot p-p^2$ in the numerators  lead finally to the result that can be  expressed by the integrals $I$ and $I_2$ . Explicitly we get:
\bea
\delta A_L(x)&=&{-c_p} \int_0^{\infty}d y y 
S_v(y)A(x) \int_0^{\infty}\frac{d \nu \rho_g(\nu)}{\nu} 
\left[y I(x,y,\nu)-\sqrt{xy}I_2(x,y,\nu)-y I(x,y,0)+\sqrt{xy}I_2(x,y,0)\right]
\nn \\
&-&\!c_p\int_0^{\infty}d y y 
B(x)S_s(y)\int_0^{\infty}\frac{d \nu \rho_g(\nu)}{\nu} 
\left[I(x,y,\nu) -\sqrt{y/x}I_2(x,y,\nu)
-I(x,y,0) +\sqrt{y/x}I_2(x,y,0)\right] \, .
\eea
Similarly the function $B_L$ can be written in the following form:
\bea
\delta B_L(x)&=&c_p \int_0^{\infty}d y y \left(  
S_v(y)B(x)+A(x)S_s(y)\right) \int_0^{\infty}d \nu
\rho_g(\nu) \frac{I(x,y,\nu)}{2}
\nn \\
&-&c_p \int_0^{\infty}d y y 
\left(S_v(y)B(x)-A(x)S_s(y)\right) 
\int_0^{\infty}d \nu \rho_g(\nu)
(y-x) \frac{I(x,y,\nu) -I(x,y,0)}{2\nu}\, .
\eea

\section{Solution of SDE in $E_T$}

Assuming regularity of  functions $S_s,S_v$ on the real axis  the quark SDE is transformed into two coupled complex integral equations which  are free of non-integrable singularities and so they are prepared for suitable numerical treatment. Beside, assuming a perturbative asymptotic ultraviolet solution, 
the SDE requires renormalization. For this purpose we use the momentum subtraction renormalization scheme, so  the SDE for unrenormalized functions $A,B$ which formally reads
\bea
B=m_o+\sum_i\delta B_i \, ; \,A=1+\sum_i\delta A_i\, ; i=T,L_\xi \, ;
\eea
are rewritten into the SDE for renormalized ones. The renormalization constant $Z_1$ is absorbed defining thus the renormalized propagator, however here we are working in $E_T$ space and a certain care is needed. First, we  avoid the mixture of different computational approaches by choosing a timelike renormalization scale. Further, we keep the renormalization constant real, thus 
only the real parts of the functions $\delta A,\delta B$ can be subtracted.
Hence the renormalization is performed as the follows:
\bea
\delta A_R(p,\mu)&=&Re \delta A(p)-Re \delta A(\mu)+i Im \delta A(p) \, ,
\nn
\\
\delta B_R(p,\mu)&=&Re \delta B(p)-Re \delta B(\mu)+i Im \delta B(p) \, ,
\eea
which leaves us with the renormalized SDE
\bea \label{brambory}
A_R(p,\mu)&=&1+\int dy \left([Re K_A(x,y)-Re K_A(\mu,y)]+ i Im K_A(x,y)\right)\, ,
\nn
\\
B_R(p,\mu)&=&m(\mu)+\int dy \left([Re K_B(x,y)-Re K_B(\mu,y)]+ i Im K_B(x,y)\right) \, ,
\eea
where, for  clarity we have explicitly indicate 
\be
\int dy K_A(x,y)=\sum \delta A_i \, ,
\ee
in order to show how the subtraction procedure works for the integral kernels.
The same is performed for similarly for the  kernel $K_B$.

As in the case of perturbation theory, the imaginary part is expected to be finite and untouched by renormalization. Clearly, this procedure maintains the hermicity of the Lagrangian.

For very low momenta the quark masses should approximately correspond to
the known values of various constituent quark models, where $M(0)\simeq \Lambda_{QCD}$ for up and down quarks. Assuming that the real part of the mass function  is  continuous  when crossing zero, this value is actually available from  Euclidean (spacelike) SDE studies: a typical estimate of the infrared mass lays in the range $250-600 MeV$, while the renormalized  mass at few GeV $m_{u,d}(2GeV)=2-8MeV$ is the standard input. Here, working in $E_T$ space instead of large, we rather choose low  renormalization scale $\mu$, concretely
\be
\mu=\Lambda_{QCD}/4
\ee
adjusting the  renormalized function  is $B(\mu)=\Lambda_{QCD}$ and 
$A(\mu)=1$.

In practice the integrals are replaced by the discrete sums on suitable grid. Setting large upper bound $e^{16}\Lambda_{QCD}$ and taking the large number $N=300-1000$ of  Gaussian mesh points shows up reasonable stability of the numeric. The functions $A$ and $B$ are separated to their real and imaginary parts and we solve resulting  four coupled integral equations   simultaneously by the method of  iterations. 
Comparing to the Euclidean spacelike case, the resulting kernel of Euclidean timelike SDE (\ref{brambory}) is not a completely smooth function  a more careful analysis is required. To speed up numeric significantly we first integrate over the integral variable $\nu$ before running the iterations.

Recently we have obtained the results for Landau gauge
$\xi=0$, where  we have achieved a good stability of our numerical solution.
In Fig. 2 we present the resulting functions obtained for 600 points
and $e^{12}\Lambda_{QCD}$ cutoff, enlarging cutoff or decreasing the number of points makes the infrared behaviour more chaotic (leaving the smooth average approximately constant).

In PT the propagator is purely real  under the threshold scale. Here, this is the main result of our presented study, the imaginary parts of the functions $B$ and $A$  are generated  below the expected perturbative threshold. The resulting  mass function becomes complex and a real pole is not present on the real axis of  square of momenta.

More interestingly, we plot the absolute values of the mass function and the inverse of renormalization function in Fig.2. The  function $|M|$ shows up the maximum at $2.3\Lambda_ {QCD}$ where it also cuts the linear function of $p$.
The phase $\phi_M$ of the mass function extracted from $M=|M|e^{i2\phi_M}$
where we got $\phi_M\simeq -25^o$ at $q=2.3\Lambda_ {QCD}$. The mass phase is a slowly varying function in the full momentum regime and it monotonously goes to small negative value in UV.

The function $A=|A|e^{i2\phi_A}$ is predominantly real, slowly varying, affecting quantitative behaviour of  the function $M$ far from the renormalization point. The results for its absolute value and phase are added to the Fig. 2 and Fig. 3 respectively. Keeping the low scale renormalization point,  it has a minimum at few $\Lambda_{QCD}$ and  it logarithmically increases in UV (the same  is also true also for spacelike regime).

In finite temperature and density QCD it is sometimes suggested, that confinement/deconfinement phenomena goes hand by hand with chiral symmetry breaking/chiral symmetric phases. In our formalism, although  there is no space for temperature definition, but the description of chiral symmetry breaking could be a QCD must. Nowadays, the lack of a reasonably precise description  of chiral symmetry breaking  is a  basic weakness of our presented $E_T$ formalism. Actually, within our setting, taking the Lagrangian mass to zero (also avoiding forbidden mass subtraction) we got the zero dynamical  mass everywhere. A bit vaguely pronounced: the kernel of the SDE is not strong enough to produce this nonperturbative effect. Without going into technical details, the phenomena of dynamical mass generation could be available   by further modeling of SDE kernel, (e.g. most naively, by further  enhancing of the gluon propagator in the infrared).  However, as we have found, the price we would pay is an unpleasant (and sometimes drastic) loss of numerical stability.

\begin{figure}
\centerline{\epsfig{figure=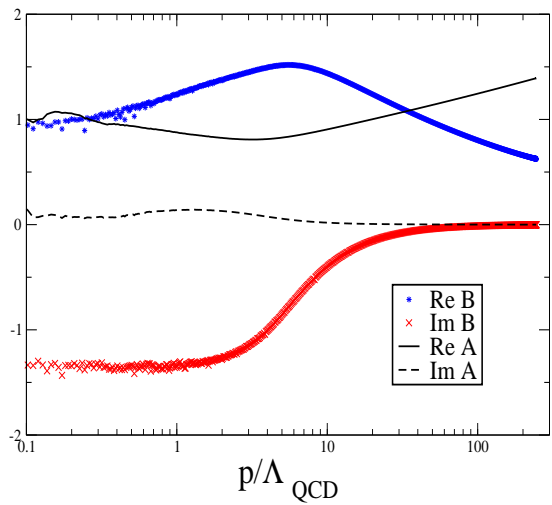,width=8truecm,height=9truecm,angle=270}}
\caption[caption]{The quark propagator function $A,B$. The upper (down) curves represent the real (imaginary) parts. The dimensionfull quantities are rescaled by QCD scale $\Lambda_{QCD}$.} \label{ETquark}
\end{figure}

\begin{figure}
\centerline{\epsfig{figure=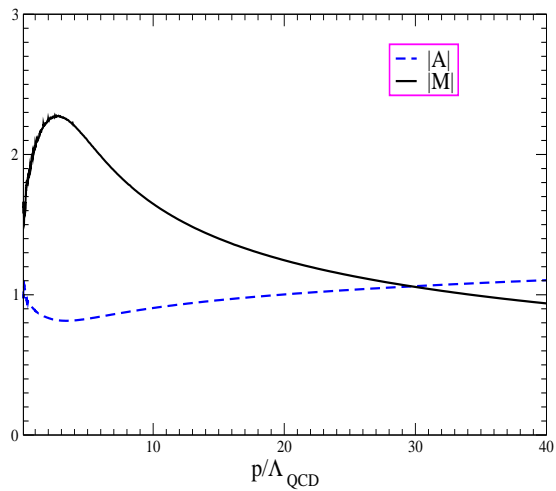,width=8truecm,height=9truecm,angle=270}}
\caption[caption]{The absolute value of quark mass function $M$ and the inverse of renormalization function- the function $||A||$ is displayed. } \label{pesek}
\end{figure}

\begin{figure}
\centerline{\epsfig{figure=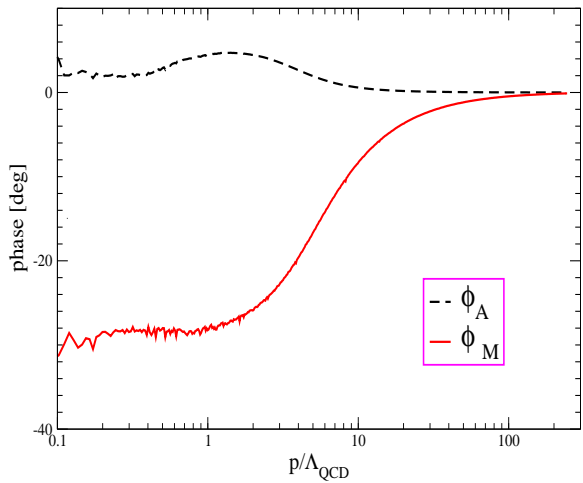,width=8truecm,height=9truecm,angle=270}}
\caption[caption]{The phase of quark mass function $M$.  The same for the function $A$.} \label{wr}
\end{figure}

\section{Summary and conclusions}

We have presented a first analysis of the quark gap equation
in the temporal Euclidean space. Given fact that 3d Wick rotated kernel 
is non-analytical function at timelike axis of momenta, 
we do not have at hand the powerful method as in the case of standard (spacelike) Euclidean formalism. Nevertheless, at this level the method really works and allows us to solve quark gap equation with a good accuracy.

 We obtain the solution with spontaneous infrared complexification of the quark mass function, as opposed to the perturbation theory, the quark mass function becomes complex from the beginning of the momentum axis. 
In  Landau gauge, $B$ is the main source of the absorptive part of the quark propagator in the infrared region, while the renormalization function appears to be  marginal for the confinement due the its small generated imaginary part. 
The absolute value of the complex mass function is enhanced at few $\Lambda_{QCD}$,  with the nonzero quark mass function phase $\phi_M\simeq 25^o$ responsible for the absence of the quark propagator  pole.

The method provides not only  a qualitative but even a quantitative 
description of propagator of confined quarks.
Following the fact that our kernel approximation is only too weak in order to produce correct chiral symmetry breaking, we can expect that the observed complexification phenomena will persist for more realistic kernels of the quark SDE.  Using the advantage of BFG, the contribution from  the longitudinal gluons to quarks selfenergy has been already fully taken into account. The product of  metric tensor with the improved quark-gluon vertex could provide the known slope of the mass function (already known form  spacelike studies). To justify our estimate explicitly, an improved study of the quark propagator with a  stable numeric is required.  

\begin{center}{\large Acknowledgment}\end{center}

We are grateful to David Emmanuel-Costa and Chitta Ranjan Das for their helps with the manuscript.

\appendix

\section{Integral $I$}

Consider the integral $I$:
\be
I(x,y,\nu)=  \int_0^{2\pi} d\theta \frac{\sin^2 \theta}{a -b \cos \theta +\ep}
\ee
where $a=x+y-\nu$ is a real number and $b=2\sqrt{xy}$ is a positive real number. Since $\nu$ is positive the integrand has a singularity in the integration range, so we keep the $\ep$ prescription of the propagator.

Making the standard substitution $t=\tan\frac{\theta}{2}$
we arrive at the following formula:

\be
I(x,y,\nu)=\frac{8}{a+b} \int_0^{\infty} 
\frac{d t \, t^2}{(1+t^2)^2(t^2-c+\ep)} \, ,
\ee
where $$c=\frac{b-a}{b+a}$$.

Performing the principal value integration one can arrive to the following result

\be
\frac{1}{\pi}I(x,y,\nu)=\frac{a}{b^2}+ \frac{b-a}{b^2}\frac{\theta(-c)}{\sqrt{-c}}-i 2 \frac{\sqrt{c}}{b} \theta{(c)}
\ee
with an integrable singularity in $\sqrt{c},(\sqrt{c})^{-1}$.

The integral 
\be
I_2(x,y,\nu)=  \int_0^{2\pi} d\theta \frac{\sin^2 \theta \, \cos \theta}{a -b \cos \theta +\ep}
\ee 
can be evaluated in a similar fashion.
 
The result is irregular at $c=0$ and regular for positive or negative $c$.
For $c>0$ it reads

\be
I_2(x,y,\nu)=\frac{\pi}{a+b}\left[\frac{1}{1+c}-\frac{8c}{(1+c)^3}
-i4\frac{(1-c)\,\sqrt{c}}{(1+c)^3}\right]
\ee

For $c<0$ we can get

\be
I_2(x,y,\nu)=\frac{\pi}{a+b}\frac{1-\sqrt{-c}}{(1+\sqrt{-c})^3}\, ,
\ee
which is again a finite function.

The special cases $\nu=0$ simplify, the functions $I(x,y,0), I_2(x,y,0)$ can be obtained by considering the appropriate limits.


\end{document}